\documentclass[conference]{IEEEtran}
\IEEEoverridecommandlockouts
\usepackage{cite}
\usepackage{amsmath,amssymb,amsfonts}
\usepackage{algorithmic}
\usepackage{graphicx}
\usepackage{textcomp}
\usepackage{xcolor}
\usepackage{balance}
\usepackage{hyperref}

\def\BibTeX{{\rm B\kern-.05em{\sc i\kern-.025em b}\kern-.08em
    T\kern-.1667em\lower.7ex\hbox{E}\kern-.125emX}}
\begin{document}

\title{Fast and Practical Strassen's Matrix Multiplication using FPGAs\\
}

\author{\IEEEauthorblockN{Afzal Ahmad, Linfeng Du, Wei Zhang}
\IEEEauthorblockA{\textit{Department of Electronics and Computer Engineering} \\
\textit{The Hong Kong University of Science and Technology, Hong Kong} \\
\{afzal.ahmad, linfeng.du\}@connect.ust.hk, wei.zhang@ust.hk
}
}
\vspace{-1in}
\maketitle


\begin{abstract}
Matrix multiplication is a cornerstone operation in a wide array of scientific fields, including machine learning and computer graphics. The standard algorithm for matrix multiplication has a complexity of $O(n^3)$ for $n\times n$ matrices. Strassen’s algorithm improves this to $O(n^{2.807})$, but its practicality is limited for small to medium matrix sizes due to the large number of additions it introduces. This paper presents a novel FPGA-based implementation of Strassen’s algorithm that achieves superior speed over an optimized General Matrix Multiply (GeMM) implementation for matrices as small as $n=256$. Our design, tested extensively on two high-performance FPGA accelerators (Alveo U50 and U280) across various data types, matches or surpasses the performance of a highly optimized baseline across a range of matrix sizes.
\end{abstract}


\maketitle

\section{Introduction}
Matrix multiplication, often referred to as General Matrix Multiply (GeMM), plays a pivotal role in a multitude of scientific domains, including but not limited to machine learning, physics simulations, and quantum computing~\cite{qiu2016going, maleki2023look, ootomo2023quantum}. Over the years, substantial resources and research have been dedicated to improving the efficiency of this crucial task~\cite{fawzi2022discovering, ahmad2020optimizing}. Specialized hardware accelerators, such as Field Programmable Gate Arrays (FPGAs) and Tensor Processing Units (TPUs), have emerged as particularly effective tools for tackling matrix multiplication problems~\cite{wang2021autosa, cong2018polysa}. Their inherent ability to parallelize repetitive computations, coupled with their significantly lower power consumption than GPUs, makes them an optimal choice for such tasks.

The traditional approach to matrix multiplication requires $O(n^3)$ operations for the multiplication of $n\times n$ matrices. This was significantly improved by Volker Strassen to $O(n^{2.807})$, achieved by employing transformations that substitute multiplications with additions, thereby unveiling a category of algorithms known as fast algorithms for matrix multiplication~\cite{strassen1969gaussian}. Despite these advancements, the implementations of Strassen’s algorithm have been limited in practicality due to two key reasons: 1) Although Strassen’s algorithm exhibits superior performance for large matrices, the overhead associated with recursion and the increased number of additions render the naive method more efficient for smaller matrices that are commonly encountered in practice~\cite{d2005using}, and 2) Strassen’s algorithm necessitates creating new submatrices during the recursion process, leading to increased memory usage~\cite{huang2016strassen}. 

Beyond these inherent limitations of the algorithm, its intricate computation pattern appears less attractive for hardware acceleration, especially compared to the traditional algorithm that benefits from a high frequency of design implementation due to its regular computation pattern. In this work, we propose a novel FPGA-based implementation of Strassen's algorithm that overcomes some of its aforementioned limitations, yielding performance benefits even for matrices as small as $n=256$. The contributions of this work are as follows\footnote{Project open-source at \url{https://github.com/afzalxo/FFGEMM}}:

\begin{itemize}
    \item We propose a fast and practical implementation of Strassen's two-level algorithm targeting FPGAs, utilizing efficient design practices yielding up to over $1.85\times$ speedup over a highly optimized GeMM baseline.
    \item Through extensive experimentation with two high-performance FPGAs, various matrix sizes, and data type bit widths, we show that the proposed kernels match or exceed the GeMM baseline's performance at a marginal increase in resource utilization.
\end{itemize}

To the best of our knowledge, this is the first Strassen-based GeMM accelerator that improves on the standard algorithm at matrix sizes as small as $n=256$.
\section{Background}
In the following section, we delve into the intricacies of implementing a high-performance standard matrix multiplication kernel. This is achieved through the utilization of systolic arrays and the application of efficient design methodologies, which include double-buffering, data streams, and shift registers. Our discussion commences with an exploration of the matrix multiplication algorithm, specifically focusing on the use of small submatrices to divide the computation. This forms the foundation of a contemporary, high-performance GeMM kernel. Subsequently, we turn our attention to Strassen’s algorithm, examining its recursive nature and the implementation of a two-level Strassen’s algorithm in our study.

\begin{figure*}[t]
\centering
\includegraphics[width=.95\linewidth]{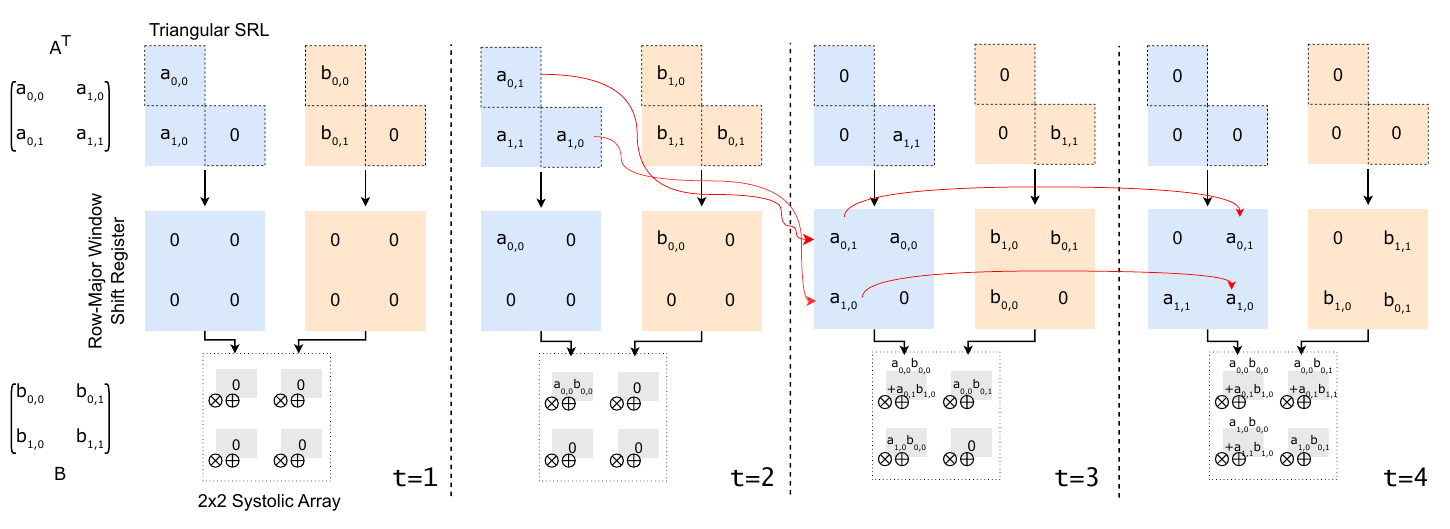}
\vspace{-0.08in}
\caption{Simplified illustration of the L1 GeMM module implemented by Vitis BLAS. The module takes $A^T$ and $B$ as inputs and utilizes shift registers and a systolic array to compute GeMM. The dotted vertical lines separate the different states of the registers as data flows through them. $t$ represents clock cycles.}
\vspace{-0.12in}
\label{fig:systolic_gemm}
\end{figure*}

\subsection{Standard Matrix Multiplication}
Consider $C=AB$ where $A$, $B$, and $C$ are $m\times k$, $k\times n$, and $m\times n$ matrices, respectively. Let $a_{i, j}$, $b_{i, j}$, and $c_{i,j}$ be the $(i,j)$ entries of $A$, $B$, and $C$, respectively. Then $c_{i,j}$, the $(i,j)$ entry of $C=AB$ is computed as the dot product of the $i$-th row of $A$ and the $j$-th column of $B$.
\begin{equation} \label{eq:matrix_cij}
    c_{i,j} = \sum_{p=1}^{k} a_{i,p} b_{p,j}
\end{equation}

Given that $A$, $B$, and $C$ can be of arbitrary size, a typical GeMM hardware accelerator employs a divide-and-conquer strategy for computing the output. This involves dividing the matrices into blocks or submatrices that can fit into the on-chip memory and then computing the GeMM result for these submatrices, accumulating the result. For instance, assuming that $m$, $k$, and $n$ are all even, we can block partition the matrices $A$, $B$, and $C$ as follows.

\begin{equation*}
A = 
\begin{bmatrix}
A_{0,0} & A_{0,1} \\
A_{1,0} & A_{1,1}
\end{bmatrix}, B = 
\begin{bmatrix}
B_{0,0} & B_{0,1} \\
B_{1,0} & B_{1,1}
\end{bmatrix}, C = 
\begin{bmatrix}
C_{0,0} & C_{0,1} \\
C_{1,0} & C_{1,1}
\end{bmatrix}
\end{equation*}

Where $A_{i,j}$ is $\frac{m}{2} \times \frac{k}{2}$, $B_{i,j}$ is $\frac{k}{2} \times \frac{n}{2}$, and $C_{i,j}$ is $\frac{m}{2} \times \frac{n}{2}$ submatrix of matrices $A$, $B$, and $C$, respectively. The output submatrices of $C$, $C_{i,j}$, can be computed using the submatrices of $A$ and $B$ in a manner similar to the scalar elements of $C$, $c_{i,j}$, as shown in Eq.~\ref{eq:matrix_cij} (i.e., $C_{i,j} = \sum_{p=1}^{k} A_{i,p} B_{p,j}$). The block partitioning of $A$, $B$, and $C$ can be performed recursively until the resulting submatrices are small enough to fit in the on-chip memory. These submatrices are then processed in parallel using a GeMM core computational unit, typically a systolic array, which is fundamental to high-performance GeMM implementations. This approach allows for efficient computation while maximizing the use of available memory and processing resources.

\subsection{Anatomy of High-Performance FPGA-based GeMM} \label{sec:intro_hpgemm}
A typical efficient FPGA-based implementation of GeMM utilizes a combination of memory and compute optimizations to achieve high performance~\cite{goto2008anatomy}. Vitis library's BLAS level 2 (L2) GeMM~\cite{vitis2023lib} is an example of a highly performant GeMM kernel. This is implemented using a variety of level 1 (L1) operations including transpose, L1 GeMM, and double buffers. In this section, we dissect the Vitis BLAS L2 GeMM kernel.

\begin{figure}[t]
\centering
\includegraphics[width=.85\linewidth]{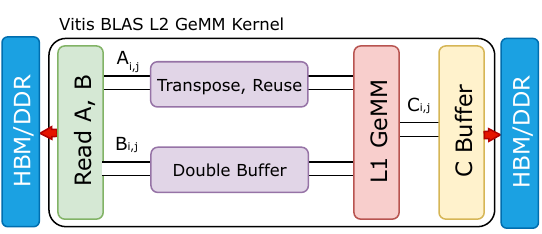}
\vspace{-0.05in}
\caption{Illustration of the L2 GeMM module implemented by Vitis BLAS. Pipes represent FIFO streams. The read module reads submatrices from external memory. Output of L1 GeMM has to be accumulated for computing the dot product between the rows and columns of $A$ and $B$. After the accumulation, the buffer contents are written to the external memory.}
\vspace{-0.2in}
\label{fig:l2_gemm_vitis}
\end{figure}

Fig.~\ref{fig:systolic_gemm} shows the Vitis BLAS L1 GeMM module which utilizes shift registers combined with a systolic array to perform GeMM. The figure presents a simplified $2\times 2$ kernel whereas practical implementations typically use larger systolic arrays (e.g., $16\times 16$) and associated registers. The module takes $A^T$ and $B$ as inputs and computes $C=(A^T)^TB$. The triangular shift register lookup tables (SRL) read $A^T$ and $B$ row by row and shift the data right along their columns in each clock cycle. The row-major window shift register loads the diagonal elements of the triangular SRL. For matrix $A^T$, the window shift register shifts the elements right along its columns, while for matrix $B$, the shift is done top to bottom along the rows of the shift register. Example flow of data from the diagonal of triangular SRL of matrix $A^T$ to the window shift register in three successive clock cycles is shown using red arrows in Fig.~\ref{fig:systolic_gemm}. The systolic array implements a 2D grid of processing elements, each containing a multiplier, accumulator, and a register. At each clock cycle, the systolic array performs element-wise multiplication of the elements in window shift registers and accumulates the result into its own registers in the corresponding rows/columns. 

\begin{figure*}[t]
\centering
\includegraphics[width=.80\linewidth]{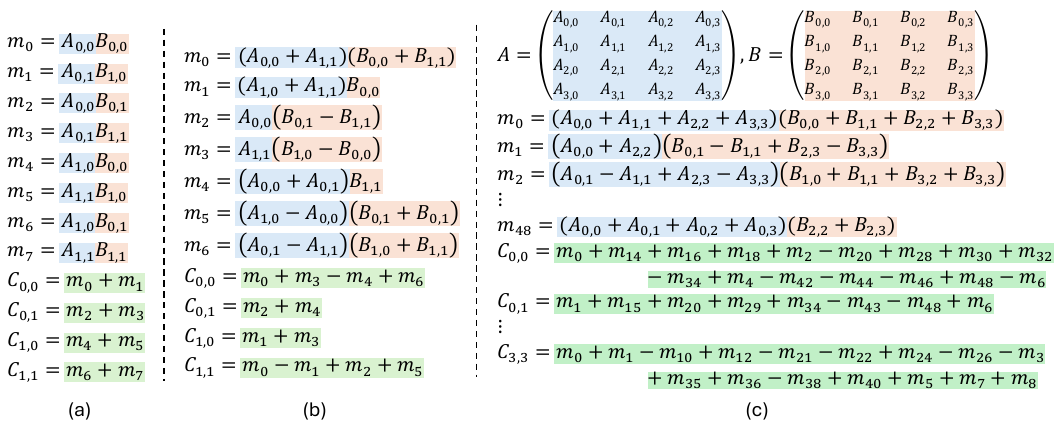}
\vspace{-0.08in}
\caption{(a) Standard GeMM and (b) Strassen's matrix multiplication algorithm for $2\times 2$ submatrices and (c) Strassen's squared algorithm for $4\times 4$ submatrices. Shown in light blue and orange colors are the left hand side (LHS) and right hand side (RHS) of the intermediate computations $m_0$ to $m_7$/$m_6$ for $2\times 2$ and $m_0$ to $m_{48}$ for $4\times 4$ case, respectively. Green color shows the accumulation of intermediate results into the output submatrices.}
\vspace{-0.12in}
\label{fig:strassens_algos}
\end{figure*}

The L2 GeMM module, shown in Fig.~\ref{fig:l2_gemm_vitis}, reads submatrices of $A$ and $B$, $A_{i,j}$ and $B_{i,j}$, of size $m'\times k'$ and $k'\times n'$, respectively. These submatrices are streamed down to the subsequent modules for consumption. The kernel utilizes a transpose operation for submatrices $A_{i,j}$, which includes reuse/buffering, and double buffering for submatrices $B_{i,j}$, followed by the L1 GeMM module. The reuse/double-buffering allow submatrices to be reused from the FIFO streams, avoiding expensive external memory transactions. For instance, $A_{0,0}$ is utilized in computation of both $C_{0,0}$ and $C_{0,1}$ for the $2\times 2$ case. The transpose module performs the transposition of $A_{0,0}$ and streams the result out twice so it can be used to compute both $C_{0,0}$ and $C_{0,1}$. Similarly, $B_{0,0}$ is utilized in computation of both $C_{0,0}$ and $C_{1,0}$. The double-buffering on submatrices of $B$ allow this reuse. Vitis BLAS extensively utilizes FIFO streams to enable efficient movement of data between the different stages of computation. By streaming data directly from one stage to the next, it reduces the need for storing intermediate results in memory, thereby saving valuable memory bandwidth and space. The implementation also makes use of standard FPGA-specific optimizations such as loop unrolling, blocking, and pipelining to further enhance performance. Given the combination of these various techniques utilized to achieve high-performance, Vitis BLAS L2 GeMM serves as a practical testbed for implementing Strassen's-based GeMM kernels.



\subsection{Strassen's Matrix Multiplication}
Fig.~\ref{fig:strassens_algos} (a) and (b) show the standard and Strassen's~\cite{strassen1969gaussian} matrix multiplication algorithms, respectively, for $2\times 2$ submatrices. Please note that Strassen's algorithm performs seven submatrix multiplications for computing the intermediate results $m_0$ to $m_6$, as opposed to eight multiplications for the standard GeMM. Computing each intermediate result, in itself, requires the multiplication of submatrices (i.e., LHS times RHS), which can be performed using another level of Strassen's algorithm, applied recursively. This recursion can be applied until the submatrices are of size $1\times 1$. However, practical implementations of Strassen's algorithm apply only a few levels of recursion until the submatrices are small enough to be efficiently multiplied using an optimized standard GeMM micro-kernel~\cite{fawzi2022discovering}. 

In this work, we utilize two levels of recursion such that the partitioned submatrices can be multiplied using the highly optimized Vitis BLAS GeMM micro-kernel. The resulting two-level algorithm for Strassen's, also known as Strassen's squared algorithm~\cite{fawzi2022discovering}, is partially shown in Fig.~\ref{fig:strassens_algos} (c) and is obtained by applying the one-level algorithm in Fig.~\ref{fig:strassens_algos} (b) recursively to a $4\times 4$ block matrix. Strassen's square algorithm is known to offer a feasible trade-off between extraneous memory requirements of overheads of multiple recursions and lack of performance from not applying the algorithm recursively~\cite{huang2016strassen,fawzi2022discovering}. Strassen's square algorithm consists of $7^2=49$ intermediate results, $m_0$ to $m_{48}$, each requiring one submatrix multiplication, as opposed to 64 submatrix multiplications for the standard GeMM, and applies to $4\times 4$ submatrices. 
\section{Related Works}
Improving the performance of matrix multiplications has received considerable interest in recent years, owing in large part to the popularity of deep neural networks (DNNs). DNN accelerators often transform convolution, an operation fundamental to computer vision, to matrix multiplication using the image to column (im2col) operation~\cite{afzal2020optimizing, ahmad2020ffconv, nurvitadhi2017can, kala2019high}. However, the high memory cost of this transformation due to the additional reshape, repack, and im2col operations becomes a hindrance in obtaining a memory-efficient implementation. Ref.~\cite{zhang2020evaluating} proposed a low-memory GeMM-based accelerator by reshaping GeMM calls to sidestep the memory bottlenecks presented by traditional GeMM-based convolution.

Works that improve Strassen's algorithm to improve its practicality utilize the system's memory hierarchy effectively to buffer/cache submatrices to achieve high performance from blocking. One such example of effective use of memory hierarchy is that of Strassen's Algorithm Reloaded~\cite{huang2016strassen}, which builds on top of an implementation of GotoBLAS~\cite{goto2008anatomy}, built using the popular BLAS-like library instantiation software (BLIS framework)~\cite{van2015blis}. The implementation treats the instructions in Strassen's algorithm as a special case of the computation $m=(X+Y)(V+W)$ and $C$ += $m$ and casts these instructions onto a highly optimized GeMM micro-kernel. This is the first implementation that yields performance benefits for matrices as small as $n=512$. However, the implementation is targeted only at CPUs. Our work aims to leverage the unique capabilities of FPGAs, such as customized high-speed on-chip buffers, to obtain speedups on Strassen's algorithm. 

Few works explore Strassen's algorithm within the context of FPGAs. Ref.~\cite{bravo2007different} performed FPGA-based implementations of the standard, Winograd~\cite{lavin2016fast,ahmad2019towards}, and Strassens~\cite{strassen1969gaussian} algorithms; however, the performance obtained is abysmal. More recently,~\cite{leon2023acceleration} utilized Strassen's-based GeMM for accelerating the fully connected layers of a DNN accelerator; however, the implementation focuses on saving DSP resources and achieves the same performance as standard GeMM at a significantly higher power consumption. Our implementation shows performance benefits for matrix sizes as small as $n=256$, utilizes similar resources as the standard GeMM implementation for \texttt{int8} data type, and achieves lower power consumption than the standard implementation. 

\section{Proposed Implementation}
\begin{figure}[t]
\centering
\includegraphics[width=.90\linewidth]{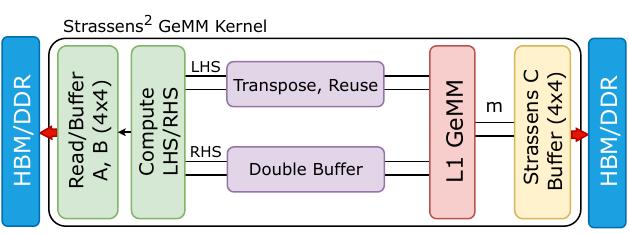}
\vspace{-0.05in}
\caption{Illustration of the Strassen$^2$ kernel. The major differences compared to the Vitis BLAS L2 GeMM (Fig.~\ref{fig:l2_gemm_vitis}) are in Read/Buffer, computation of LHS/RHS, and the output buffering.}
\vspace{-0.18in}
\label{fig:strassens_sq_block_diagram}
\end{figure}

This section outlines our proposed Strassen's squared GeMM kernel implementation, which utilizes several design optimizations beyond the traditional GeMM implementation. Specifically, we outline input buffering/reuse, output buffering, computation of transformed inputs, and various other optimizations. Figure.~\ref{fig:strassens_sq_block_diagram} shows an illustration of our implemented kernel. Our implementation is plug-compatible with the Vitis BLAS L2 GeMM. Hence, the host program does not change. The host serves to allocate input/output buffers, copy the input matrices to the devices' HBM/DDR global memory, and retrieve the output from the device's output memory space. 

\subsection{Input Buffering/Reuse} \label{proposed:input_buffering}
The most striking difference between the traditional GeMM and Strassen's algorithm is how the inputs are utilized. While traditional GeMM directly multiplies two input submatrices from $A$ and $B$, Strassen's algorithm needs to compute input transformations using multiple submatrices of both input matrices (See for example $m_0$ in Fig.~\ref{fig:strassens_algos} (c) which requires four submatrices from each of $A$ and $B$). Therefore, every intermediate computation ($m_0$ to $m_{48}$) would require loading multiple submatrices of both $A$ and $B$ from the external memory onto the on-chip memory. If these submatrices are not already present in the on-chip memory, the overheads of frequent off-chip memory transactions will far exceed the performance benefits from the reduction in number of multiplications, given the same submatrices are utilized by multiple intermediate computations (e.g., Fig.~\ref{fig:strassens_algos} (c) shows submatrix $A_{0,0}$ is required by $m_0$, $m_1$, $m_{48}$, and many others). We address this challenge of extraneous external memory access burden by allocating on-chip block-RAM buffers for $4\times 4$ submatrices of $A$ and $B$ to store all 16 submatrices needed to compute $m_0 ... m_{48}$. Each of the $4\times 4$ submatrices is of size $m'\times k'$ for $A$ and $k' \times n'$ for $B$. 

The kernel begins execution by loading the $4\times 4$ blocks of both $A$ and $B$ from the external memory (HBM/DDR) onto the on-chip BRAM. Given that these blocks are loaded in mostly a contiguous fashion, external memory requests utilize bursts to hide the access latency of data transfer for contiguous, sequential data, allowing us to maximize the HBM/DDR memory bandwidth. We utilize bursts of length $4\times k'$ and $4\times n'$ for submatrices of $A$ and $B$, respectively. Buffering the submatrices allows us to reuse them for every intermediate computation $m_0, ..., m_{48}$, without ever having to perform another external memory transaction for the entirety of the computation of the current block multiplication. 

\begin{figure*}[t!]
\centering
\includegraphics[width=.80\linewidth]{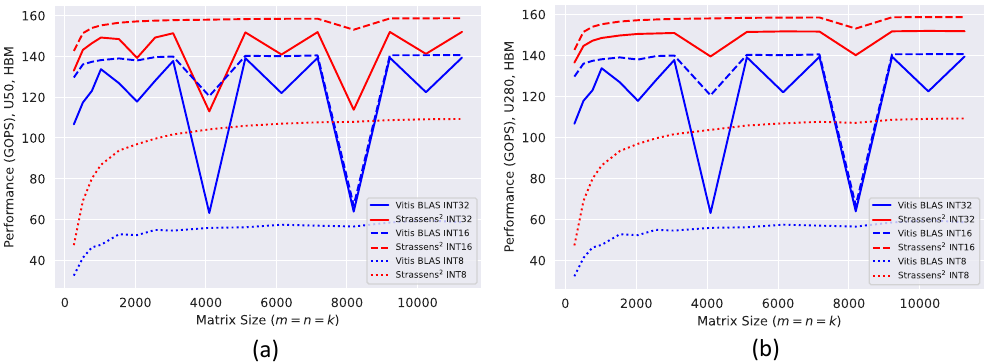}
\vspace{-0.12in}
\caption{Comparison of measured performance of Strassen square kernel against Vitis BLAS GeMM for different data types and matrix sizes, on (a) Alveo U50 and (b) Alveo U280 FPGAs.}
\vspace{-0.2in}
\label{fig:plot_perf_u50_u280}
\end{figure*}

\subsection{Computing LHS and RHS}
Having the $4\times 4$ submatrices in the on-chip memory, the buffered blocks can be used to compute the LHS and RHS (marked in blue and orange on Fig.~\ref{fig:strassens_algos} (c)) for each of the intermediate instructions without performing any external memory transactions. Strassen's squared algorithm utilizes either four, two, or one operand on LHS and RHS of the 49 intermediate instructions, requiring the same number of additions/subtractions. Hence, we implement three modules corresponding to the number of operands. These modules simply read different submatrices from the input BRAM buffers and compute the LHS and RHS, placing the result in FIFO streams. The modules operate on submatrices of size $m'\times k'$ for submatrices of $A$ and $k'\times n'$ for submatrices of $B$. The LHS and RHS results are computed in parallel since they require reading from separate buffers. Standard FPGA design optimizations such as pipelining and unrolling are utilized to process multiple memory words in parallel.

\subsection{Computing $m_0, ..., m_{48}$}
The FIFO streams for LHS and RHS are consumed by the subsequent GeMM module. We utilize Vitis BLAS L2 GeMM's Transpose/Reuse, Double Buffer, and L1 GeMM modules for this purpose, allowing a fair comparison and making our kernel plug-compatible. We refer to this section of Vitis BLAS as the GeMM micro-kernel. We make $49$ calls to the micro-kernel per $4\times 4$ block multiplication. The micro-kernel consumes the LHS and RHS streams and computes the intermediate results, streaming them down in another FIFO. For details of this micro-kernel, please see Section.~\ref{sec:intro_hpgemm}.

The submatrices resulting from the intermediate results (i.e., $m_0$ to $m_{48}$) appear to require additional memory for storage. This is one of the reasons cited for the impracticality of Strassen's algorithm~\cite{huang2016strassen}. However, a closer inspection of the algorithm reveals that multiple intermediate results need not be kept in memory. Instead, whenever an intermediate result is computed, it can directly be accumulated into the output submatrices that need it (e.g., when $m_0$ finishes computation, it can be accumulated into output buffers corresponding to $C_{0,0}, C_{3,3}$ and so on). This allows the space/FIFO-depth utilized by $m_0$ to be freed and reused for computing $m_1$. 

\subsection{Output Buffers} \label{proposed:output_buffers}
The outputs $C_{0,0}$ to $C_{3,3}$ represent the $4\times 4$ output submatrices, each of size $m'\times n'$. The intermediate results $m_0$ to $m_{48}$ arrive in a FIFO stream and need to be accumulated to obtain the output submatrices. Some intermediate results need to be accumulated into multiple output submatrices (e.g., $m_0$ needs to be accumulated into $C_{0,0}$, $C_{3,3}$, and many others). The accumulation of the intermediate results into the outputs requires both reading from and writing to the outputs $C_{0,0}$ to $C_{3,3}$. Hence, it is infeasible to perform this operation directly on external memory due to the high number of transactions required. We create a $4\times 4$ output submatrix buffer using on-chip BRAMs for accumulating the intermediate results arriving via the FIFO stream. The accumulations are performed in parallel and pipelined fashion into multiple submatrices of $C$. This alleviates the high cost of this operation in Strassen's algorithm relative to a standard GeMM implementation, which only requires accumulating each output into a single submatrix (e.g., In Fig.~\ref{fig:strassens_algos} (a) $m_0$ is consumed only by $C_{0,0}$ while in (c) $m_0$ is consumed by $C_{0,0}, C_{3,3}$, and many others). Once all 49 intermediate results are accumulated into the output buffers $C_{0,0}$ to $C_{3,3}$ according to the instructions in Strassen's squared algorithm, the buffer is written to the global memory. These memory writes also utilize bursts of length $4\times n'$.

\subsection{The Outer Loops}
Since computing a single output submatrix $C_{i,j}$ requires many submatrix multiplications ($\sum_{p=1}^{k} A_{i,p} B_{p,j}$), the process described in subsections~\ref{proposed:input_buffering} to~\ref{proposed:output_buffers} is nested inside three loops. The innermost loop iterates over the submatrices in columns/rows of matrices A/B. The inner loop iterates over the submatrices in columns of matrix $C$, while the outermost loop iterates over the submatrices in rows of matrix $C$.

The entire flow described in this section is part of a dataflow region, implemented as tasks that run in parallel. While one task is consuming a stream, the previous task places more data into the stream to be processed so that the tasks can be executed in a pipelined fashion. 

\subsection{OpenCL Host}
Our kernel is plug-compatible with the Vitis BLAS GeMM. This means that we utilize the same interfaces and global memory space as Vitis BLAS, allowing for a fair comparison. After the kernel finishes execution, the host reads the result of the GeMM computation from the device's global memory and compares the kernel's output with the result of the host-side execution of GeMM to ensure the correctness of results.

\section{Experiments and Results}
We implemented both the baseline Vitis BLAS L2 GeMM and our Strassen's squared kernel for two FPGA platforms, Alveo U50 and U280. We set the kernel's target frequency to $275$ MHz and implement the kernel for three data types: \texttt{int32}, \texttt{int16}, and \texttt{int8}. We set $m'$, $k'$, and $n'$ to $64$ so the $4\times 4$ submatrices fit into the on-chip BRAMs. All implementations meet the timing constraints. The GeMM micro-kernel, utilized by both ours and the baseline implementations, uses a systolic array of size $16\times 16$. We set the parallelism factors such that the designs fit onto a single die of each of the two FPGAs. To ensure fairness of comparison, we utilize the same parallelism factors and memory banks for both designs. 

\begin{figure}[t]
\centering
\includegraphics[width=.85\linewidth]{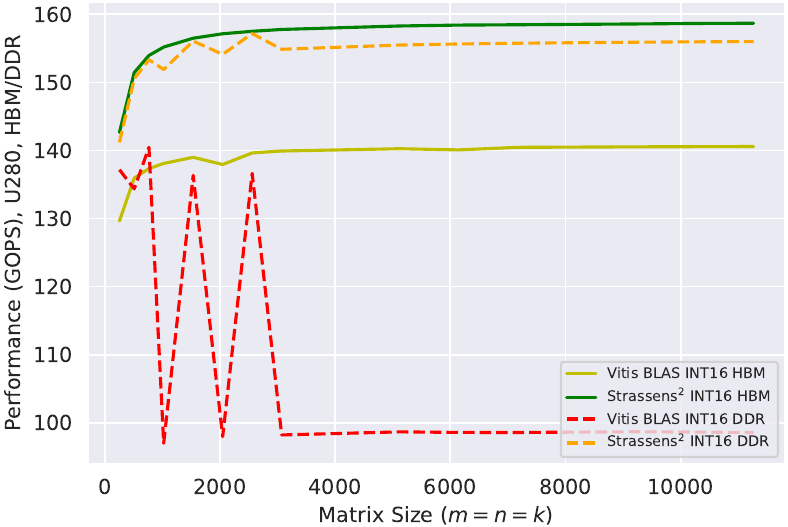}
\vspace{-0.12in}
\caption{Comparison of measured performance of Strassen squared kernel and Vitis BLAS GeMM kernel on DDR and HBM memory interfaces on \texttt{int16}.}
\vspace{-0.16in}
\label{fig:plot_perf_u280_ddr_hbm}
\end{figure}

We measure the kernel performance on the FPGA accelerators using the number of clock cycles passed from the start to finish of the kernel execution. This is measured using a hardware clock cycle counter that runs in parallel to the GeMM kernel. The kernel's runtime $t$ is computed using $t=clock\_cycles/frequency$. This is then used to compute the kernel's performance in giga operations per second (GOPS):

\begin{equation}
    GOPS = \frac{2mkn}{t}\cdot 10^{-9}
\end{equation}

We obtain this kernel performance at various square matrix sizes, with the smallest being $m=k=n=256$, and for each of the three data types and the two FPGA accelerators. Since the U280 FPGA offers both DDR and HBM memory interfaces, we implement designs using both these interfaces to compare their performance differences. 

\subsection{Results and Comparison}
Fig.~\ref{fig:plot_perf_u50_u280} (a) and (b) show the measured performance in GOPS against matrix size for Alveo U50 and U280 HBM, respectively. Our \texttt{int16} design achieves the highest performance, reaching a peak of $158.8$ GOPS. This is around 14\% higher than Vitis BLAS's peak performance on this datatype. Perhaps more prominent is the result on \texttt{int8} datatype on which our kernel achieves around $1.85\times$ better performance. This is likely due to the better memory bandwidth utilization since loading $4\times 4$ submatrices allows us to use the external memory bandwidth more efficiently. The baseline kernel reaches peak performance of only $59.1$ GOPS at this data type. This is likely because of the short width and depth of the memory transactions resulting from the small bitwidth of the data type. 

The results for the two FPGAs are fairly similar while using the HBM memory interface, aside from the \texttt{int32} implementation, which performs more consistently on the U280 FPGA. There are sudden and severe performance drops at matrix sizes that are multiples of $4096$. These performance drops are significantly more severe in the baseline implementation. We conjecture that these are due to the utilization of multiple HBM memory banks on the same kernel interface. When the matrix sizes exceed a certain threshold, the kernel has to search for the appropriate HBM bank as these banks are low capacity (e.g., a single \texttt{int32} matrix of size $n=8192$ requires $256$ MiB of storage, and we store $3$ matrices concurrently; some matrix sizes exceed the capacity of a single HBM bank). Furthermore, these performance drops are not exhibited when using the DDR interface since a single DDR memory bank is much larger than an HBM memory bank (HBM banks are $256$ MiB each while a DDR bank is $16$ GiB). 

\subsection{HBM vs. DDR}
Fig.~\ref{fig:plot_perf_u280_ddr_hbm} shows the performance of baseline and Strassen's squared kernels in \texttt{int16} for both HBM and DDR memory interfaces (performance outliers removed). Unsurprisingly, HBM-based implementations always perform better compared to their DDR-based counterparts. Our Strassen's-based kernel achieves similar performance on the two memory interfaces at around $155$ GOPS. On the other hand, the baseline kernel achieves around $140$ GOPS in the steady state using HBM and around $100$ GOPS using DDR. 

\setlength{\tabcolsep}{2pt} 
\begin{table}
  \begin{minipage}{\linewidth}
    \centering
    \caption{Resource/Power Comparison of Strassens$^2$ kernel agianst Vitis BLAS L2 GeMM on Alveo U50.}
     \vspace{-0.04in}
    \resizebox{\linewidth}{!}{
	\begin{tabular}{l|ccc|ccc}
		\hline
            Implementation & \multicolumn{3}{c|}{Vitis BLAS GeMM} & \multicolumn{3}{c}{Strassens$^2$ GeMM \textbf{(Ours)}} \\\hline 
		  DataType & \texttt{int32} & \texttt{int16} & \texttt{int8} & \texttt{int32} & \texttt{int16} & \texttt{int8}\\\hline
            Achieved Freq (MHz) & 275 & 275 & 275 & 275 & 275 & 275 \\
            LUT & 112k (25\%) & 71k (16\%) & 99k (22\%) & 155k (35\%) & 105k (24\%) & 102k (23\%)\\
            FF & 153k (17\%) & 56k (6\%) & 45k (5\%) & 167k (19\%) & 61k (7\%) & 40k (4\%)\\
            DSP & 784 (26\%) & 272 (9\%) & 272 (9\%) & 784 (26\%) & 272 (9\%) & 272 (9\%)\\
            BRAM 18K & 64 (4\%) & 48 (3\%) & 8 (0\%) & 336 (25\%) & 208 (15\%) & 104 (7\%)\\
            Peak GOPS & 139.4 & 140.6 & 59.6 & 151.9 & 158.7 & 110.2 \\
            Dynamic Power (W) & 16.7 & 15.1 & 14.8 & 11.8 & 10.3 & 10.2\\
            \hline
	\end{tabular}
 }
    \label{tab:resource_power_u50}
     \vspace{-0.20in}
  \end{minipage}
\end{table}

\subsection{Resource/Power Consumption Comparison}
Table.~\ref{tab:resource_power_u50} shows the resource and power consumption of both the Vitis BLAS GeMM and our Strassens$^2$ kernel on the Alveo U50 FPGA on all three data types. Resource utilization percentage consumption is relative to a single die/SLR, whereas the FPGA platform has two dies. Dynamic power is obtained from the post-implementation report, which uses analysis of the implemented netlist to estimate the power consumption of the design. The resource utilization of our designs is marginally higher than the baseline. Of particular significance to our design is the aggressive use of buffers for inputs and outputs, which leads to up to $21\%$ higher utilization of BRAMs in the Strassen$^2$ kernel compared to the baseline. Nevertheless, the bottleneck resource in nearly all designs is logic resources rather than BRAMs. Our design for \texttt{int8} consumes nearly similar resources as the baseline while yielding $1.85\times$ better GOPS and consuming around 38\% less power. This means that Strassen's-based kernels can be particularly resource/power efficient for low-precision data types that FPGAs are known to be friendly towards. For high-precision data types, the overheads of additional logic (such as multiplexers for addressing more BRAMs than the baseline) increases the resource utilization drastically.

\section{Conclusion}
This paper presented a novel FPGA-based implementation of Strassen's algorithm for matrix multiplication. Aiming to obtain speedups relative to traditional implementations of GeMM by exploiting the reduction in number of calls to the GeMM micro-kernel, we showed, for the first time, that a Strassen's-based implementation can exceed the performance of traditional GeMM for practically sized matrices as small as $n=256$. Our \texttt{int8} implementation achieved over $1.85\times$ the performance of highly optimized Vitis BLAS GeMM at a similar resource utilization, showcasing the ability of Strassen's-based GeMM implementations for FPGA-friendly, low-precision data-types.

\balance
\bibliographystyle{ieeetr}
\bibliography{biblio}

\end{document}